\documentclass[aps,prb, twocolumn, superscriptaddress]{revtex4}

\usepackage{graphicx}
\usepackage{tabularx}
\usepackage{dcolumn}
\usepackage{bm}
 \usepackage{amsmath, amsthm, amssymb, subeqnarray}

\begin{document}
\indent

\title{Magnetic Fluctuations and Correlations in MnSi \\ Evidence for a Skyrmion Spin Liquid Phase}

\author{C. Pappas}
\affiliation{Delft University of Technology, Faculty of Applied Sciences, Mekelweg 15, 2629JB Delft, The Netherlands}
\affiliation{Helmholtz Center Berlin for Materials and Energy, Glienickerstr. 100, 14109 Berlin, Germany}
\email{c.pappas@tudelft.nl}
\author{ E. Leli\`evre-Berna}
\affiliation{ Institut Laue Langevin, 6, Rue Jules Horowitz, 38042 Grenoble, France}
\author{ P. Bentley}
\affiliation{ Institut Laue Langevin, 6, Rue Jules Horowitz, 38042 Grenoble, France}
\author{P. Falus}
\affiliation{ Institut Laue Langevin, 6, Rue Jules Horowitz, 38042 Grenoble, France}
\author{ P. Fouquet}
\affiliation{ Institut Laue Langevin, 6, Rue Jules Horowitz, 38042 Grenoble, France}
\author{ B. Farago}
\affiliation{ Institut Laue Langevin, 6, Rue Jules Horowitz, 38042 Grenoble, France}

\date{\today}

\begin{abstract}
We present a comprehensive analysis of high resolution neutron scattering data involving Neutron Spin Echo spectroscopy and Spherical Polarimetry which confirm the first order nature of the helical transition and reveal the existence of  a new spin liquid skyrmion phase. Similar to the blue phases of liquid crystals this phase appears in a very narrow temperature range between the low temperature helical and the high temperature paramagnetic phases. 
\end{abstract}
\maketitle

\section{Introduction}

MnSi  is one of the most investigated chiral magnets and, at least in theory, a very simple realization of chiral magnetism. The Ginzburg-Landau Hamiltonian contains three hierarchically ordered interaction terms with well separated energy scales\cite{Bak}. It is therefore possible to distinguish between the different contributions to the ground state. The strongest ferromagnetic exchange interaction term fixes the spins at the longest range. The weaker Dzyaloshinskii-Moriya (DM) term arises from the lack of inversion symmetry of the B20 lattice structure and rotates the spins at the intermediate scales. And the weakest anisotropy term fixes the directions of the spins on the crystallographic lattice. The magnetically ordered state below T$_{C}\sim$ 29 K is a single domain left handed helix with a period of $\sim$175 \AA$\,$. All magnetic moments are perpendicular to the propagation vector $\vec{\tau}$ that points along the $<$111$>$ crystallographic directions \cite{Shirane, Ishida}.  In the helical phase small angle neutron scattering shows well defined Bragg peaks of the same intensity at all equivalent magnetic reflections $|\vec{\tau}_{111}|$, with $\tau$= $|\vec{\tau}_{111}|$=0.036 \AA$^{-1}$. 
\\ \indent
MnSi is a weak itinerant magnet with an ordered magnetic moment of only 0.4 $\mu_B$, a fraction of the effective magnetic moment of 1.4  $\mu_B$ determined in the paramagnetic phase\cite{ordered_moment}. The strong magnetic fluctuations, which are due to the weak itinerant magnetism and the vicinity of a magnetic instability, exist not only  above  T$_{C}$ \cite{Ishikawa} but also persist in the low temperature ordered phase\cite{Ziebeck}. The magnetic fluctuations lead to an enhanced effective electron mass, a broad specific heat maximum  reminiscent of the specific heat of spin liquids, frustrated magnets or spin glasses\cite{StishovJPhysC} and broad features on thermal expansion and ultrasound measurements, which almost completely mask the helical transition\cite{Petrova}. 
\\ \indent
One intriguing feature of MnSi is that magnetic correlations above T$_C$ appear not only around the positions in reciprocal space of the helical order but spread homogeneously over the whole surface of a sphere with radius  $\tau$ emerging as a powder diffraction like ring on the two-dimensional small angle neutron scattering spectra\cite{Grigoriev}. If the neutron beam is polarized the rings reduce to half-moons due to the interaction between polarized neutrons with the helical correlations as explained below. 
\\ \indent
This unconventional  feature occurs in a limited temperature range above T$_C$ and is reminiscent of scattering patterns from cholesteric liquid crystals suggesting the existence of similar textures also in magnets. However, the chiral molecules of liquid crystals are rods whereas magnetic moments are vectors. This additional topological constraint leads to the formation of domain walls rendering $\pi$ disclinations, i.e. the low order line defects commonly found in liquid crystals, energetically unfavorable in magnets\cite{Wright}. On the other hand, domain wall formation is minimized by higher order skyrmion-like 2$\pi$ disclinations. 
			\begin{figure*}
			\includegraphics[width=140mm]{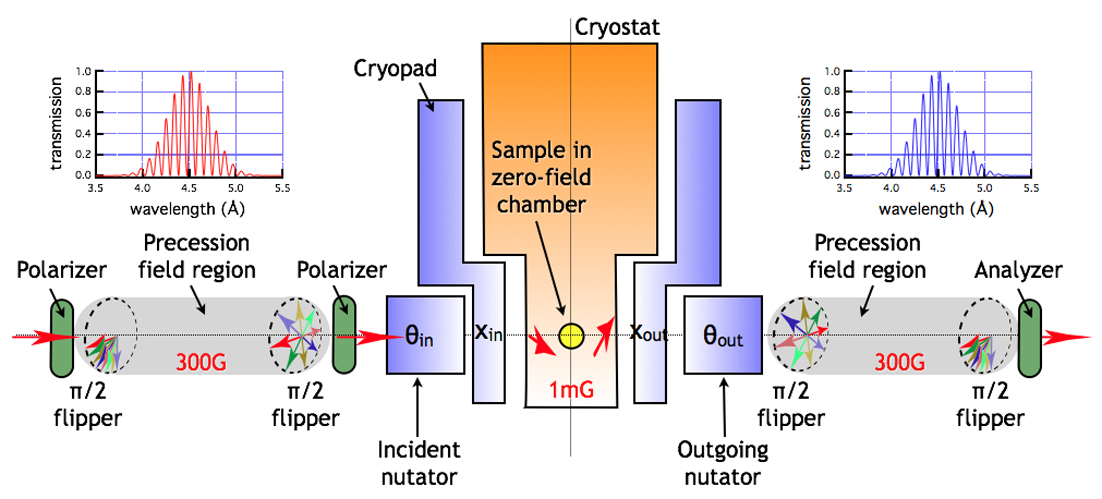}
			\caption{Principle of the Polarimetric Neutron Spin Echo technique. The neutron wavelength information is encoded in the first precession arm between two $\pi/2$ flippers. The beam is re-polarised before entering the zero-field region of Cryopad. The incident and scattered polarisation vectors are set by Cryopad. After scattering, the second precession arm, between two additional $\pi/2$ flippers, encodes again the wavelength and the echo is measured after the analyser.}
			\label{polnse_schema}
			\end{figure*}
\\ \indent		
Skyrmions were  introduced in the early 60s by Skyrme \cite{Skyrme} to bridge the gap between waves and particles in the particle-wave duality description. The existence of these soliton-like quasiparticles was alleged in semiconductors under high magnetic fields \cite{GaAS} and their topology corresponds to that of  the blue phases of liquid crystals \cite{blue_phases}.  Periodic skyrmion lattices have recently been identified under a magnetic field in the A-phase of MnSi\cite{Skyrmion_lattice} or directly seen by Lorentz transmission electron microscopy in Fe$_{0.5}$Co$_{0.5}$Si\cite{Lorenz_TEM} or FeGe\cite{FeGe_Lorenz_TEM}.
\\ \indent
It has been suggested that the unconventional  diffuse scattering above T$_C$  reflects the existence of skyrmion-like textures, which would form spontaneously without any structural defects or an external magnetic field\cite{Skyrmion} stabilized by spin stiffness or higher order terms \cite{Binz1, Fischer}.  This novel ground state of condensed matter would result from the isotropically twisted spin liquid phase and as in liquid crystals it would occur at a very restricted temperature range just above the long range ordered helical phase. In the following we present high resolution polarized neutron scattering experimental results, which strongly support this hypothesis.
			\begin{figure}
			\includegraphics[width=85mm]{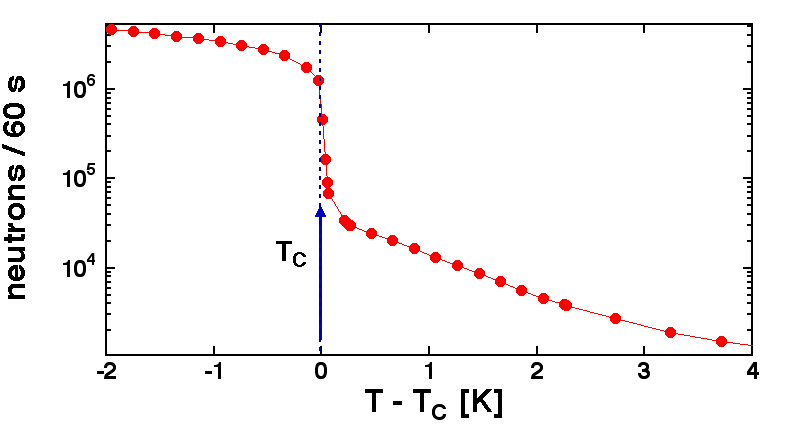}
			\caption{Temperature dependence of the logarithm of the intensity at the position of the helical peak ($\vec{\tau}_{111}$= 0.036~\AA$^{-1}$). A intensity jump of almost one order of magnitude defines the helical transition T$_C$.}
			\label{intensity}
			\end{figure}
\section{Experimental Method}
 \indent
MnSi crystallizes in the cubic $P2_13$ ($T4$) structure with the Mn atoms occupying the 4a site for $x = 0.138$. The lack of inversion symmetry in the crystallographic structure is at the origin of the DM interaction, the chiral magnetism and the helical magnetic structure. The experiments were done on a well characterized \cite{Grigoriev, Okorokov} single crystalline sample with a thickness of 2 mm and a diameter of 20 mm, cut from a large single crystal grown at Ames Laboratory. The lattice constant was 4.558 \AA. The sample was oriented with a $<$110$>$ direction vertical, so that four 111 reflections and two 110 reflections were accessible in the horizontal scattering plane. 
\\ \indent
The polarized neutron scattering experiments were carried out with the Neutron Spin Echo spectrometers IN11 and IN15 at the Institut Laue Langevin, Grenoble, France. On IN15 the wavelength was 8~\AA \ and on IN11  6.5~\AA. At both instruments the incoming neutron beam had a monochromatization of 15\% FWHM. The XY position sensitive detectors of both  spectrometers covered an angular range of $\sim$3x3 deg and  the resolution in momentum transfer $Q$  was $\sim$ 0.005\;\AA$^{-1}$ FWHM. On IN11 the $Q$ dependence of the relaxation and the intensity was analyzed by taking slices of the detector at constant distance from the 000 point and also from the surface of the sphere with radius  $\tau$  leading to $-0.005~ \mathrm{\AA}^{-1} \leq q = |\vec{Q}|-\tau \leq 0.018~ \mathrm{\AA}^{-1}$. The measurements were carried out at one of the four equivalent magnetic reflections  000$+\vec{\tau}_{111}$.  On IN11 it was also possible to measure at a 000$+\vec{\tau}_{110}$ point, where the correlations do not develop to a Bragg peak. 
\\ \indent
Fig.~\ref{intensity} shows on a log-lin scale the temperature dependence of the neutron intensity at $\vec{\tau}_{111}$. The helimagnetic phase transition is marked by an intensity jump of more than one order of magnitude within less than 0.1 deg. The data above and below T$_C$ follow power laws similar to those found previously \cite{Grigoriev} and which were at the origin of a long standing controversy on the order of the transition. These power laws, however, do not account for the intensity jump, which confirms the first order phase transition seen by specific heat\cite{StishovPRB}. 
\\ \indent
The intensity jump defines very accurately the transition temperature is T$_C$ = 29.05 $\pm$ 0.05~K.  These data were collected  on IN15. A different cryostat and thermometer were used on IN11 and the jump was observed at T$_C$ = 28.6 $\pm$0.05 K. In order to compare the data sets the results will be plotted against T-T$_C$ or the reduced temperature $\epsilon = $(T-T$_C$)/T$_C$. 
\\ \indent
The helical Bragg peaks are the fingerprint of the helical phase. They are elastic (energy transfer $\hbar\omega=0$, in the following we will often use $\omega$ to also designate the energy transfer without explicitly using $\hbar$) and has a Gaussian lineshape. In contrast, the fluctuating paramagnetic phase above T$_C$ has finite correlations and the scattering function $S(Q, \omega)$ is a superposition of Lorentzians\cite{Ishikawa, Grigoriev}. In the quasi-elastic limit, where the energy transfer is much smaller than the energies of the incoming beam and of the sample, the neutron scattering cross section becomes:
	\begin{equation}
	\frac{d^2\sigma}{d Q \, d E} \,  \propto \, S(Q, \omega) \propto \frac{C}{q^2+\kappa^2} \; \frac{\Gamma}{\Gamma^2 + \omega^2} 
	\label{s_q_omega}
	\end{equation}
with $C$ the Curie constant,  $\kappa=1/\xi$, where $\xi$ is the characteristic correlation length, and $\Gamma$ the HWHM energy linewidth. $S(Q,\omega)$ is therefore the product of the static structure factor, which has the Ornstein-Zernike (OZ) form  :
	\begin{equation}
	S(Q)= \int{S(Q, \omega)\; d\omega} \propto  \frac{C}{q^2+\kappa^2}
	\label{static}
	\end{equation}
and of the dynamic structure factor:
	\begin{equation}
	s(Q, \omega)=\frac{S(Q,\omega)}{S(Q)} \propto \frac{\Gamma}{\Gamma^2 + \omega^2}
	\label{dynamic}
	\end{equation}
\\ \indent
The dynamic measurements were performed by Neutron Spin Echo spectroscopy, which uses the Larmor precession of the neutron spin in a magnetic field as a clock to measure with very high accuracy the difference in neutron velocities before and after the scattering process at the sample. The  changes in the neutron energy due to inelastic scattering affect the neutron velocity and consequently the amplitude of the Larmor precessions. In this way, the energy transfer is measured directly by circumventing the intensity resolution limitations of the Liouville theorem. For this reason NSE reaches very high resolutions while maintaining the high intensity advantage of a beam that is only 10-20\% monochromatic. The highest energy resolution in neutron scattering is presently reached by IN15, which accesses energies as low as some neV corresponding to motions with characteristic times reaching the $\mu s$.
\\ \indent
At the quasi-elastic limit,  which is valid in most NSE experiments, the amplitude of the NSE Larmor precessions is directly proportional to the intermediate scattering function $I(Q,t)$, the Fourier transformation of $s(Q,\omega)$\cite{FeriNSE}. The comparison between NSE and inelastic neutron scattering is therefore straightforward through a Fourier transformation.
	\begin{equation}
	I(Q, t)=\frac{\int{S(Q,\omega) \; cos(\omega \,t) \; d\omega}}{\int{S(Q, \omega)\; d\omega}} = \frac{\Re(S(Q,t))}{S(Q)}
	\label{inter}
	\end{equation}
\indent 
Consequently, the Lorentz function of eq.~\ref{dynamic} Fourier transforms to the exponential NSE decay $\exp(-t/t_0)$ with $t_0=1/\Gamma$ leading to $t_0[ns]=0.658/\Gamma$[$\mu$eV]. With this relation it will be possible to compare our NSE results with the triple axis neutron spectroscopy (TAS) measurements from the literature\cite{Ishikawa, Roessli}. 
\\ \indent
This so-called Larmor labeling  requires polarized neutrons and  some polarization analysis features are  integral part of NSE\cite{MEZEI_Murani}. The pulse sequence, $\pi/2$ flip-precession-$\pi$ flip (at the sample)-precession-$\pi/2$ flip, is that of the classical Hahn sequence of NMR SE.  ``Normal'' paramagnetic scattering acts as a $\pi$  flipper and gives an echo without the otherwise obligatory $\pi$ flipper, which leads to a straightforward and unambiguous separation of the magnetic and nuclear signals. On the other hand, due to Larmor precessions, the neutron beam is depolarized at the sample with all neutron spins evenly distributed  in the precession plane. For this reason, in the presence of  chiral and/or nuclear-magnetic interference terms the analysis of the experimental results is complex. The way out is the Polarimetric Neutron Spin Echo set-up, a variant of Intensity Modulated NSE\cite{IMNSE},  which combines the precession field areas required for Neutron Spin Echo spectroscopy with Cryopad \cite{POLNSE}. As shown schematically by fig.~\ref{polnse_schema}, the precessions are stopped before the sample by an additional $\pi/2$ flipper. The neutron beam is then re-polarized with a compact polarizer to allow full analysis of the scattered polarization vector with the zero field chamber of Cryopad. The precessions resume at the second branch of the spectrometer after a second additional  $\pi/2$ flipper and the echo signal is recovered at the neutron detector after the analyzer.  
\\ \indent
This set-up is now implemented  at IN15 and can be used to observe chiral fluctuations with unprecedented accuracy and resolution both in energy (time) and momentum transfer (space). The Polarimetric NSE and spherical polarimetry measurements were performed with a third generation Cryopad, a sensitive zero field polarimeter which controls the polarization of the incoming and the scattered beams\cite{Cryopad1, Cryopad2}. A combination of $\mu$-metal  and Meissner shields reduces the residual magnetic field at the sample position down to  $\sim$0.1~$\mu$T and controls the direction of the polarization vectors with an accuracy of better than  1~deg.  The  beam polarization was 96\% corresponding to a flipping ratio of 45.
\section{Polarized Neutron Formalism - Spherical Polarimetry} \label{pol}
			\begin{figure}
			\includegraphics[width=85mm]{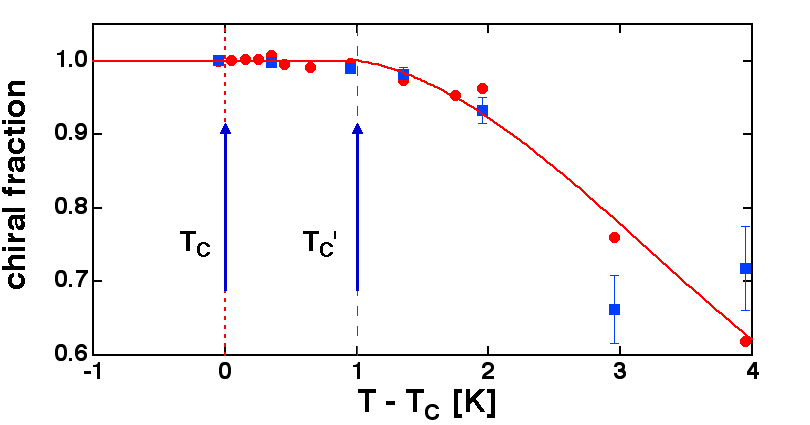}
			\caption{Temperature dependence of the chiral fraction determined by neutron polarimetry as explained in the text. }
			\label{chiral_fraction}
			\end{figure}
\indent
Before presenting in detail the experimental results we will introduce the interaction between a polarized neutron beam and the magnetic helix of MnSi following the formalism developed almost simultaneously by Blume\cite{Blume} and Maleyev\cite{Maleyev} in 1963. The helix is described by two orthogonal vectors $\vec{s_1}$ and $\vec{s_2}$, which have  the same amplitude and are  perpendicular to the helix propagation vector $\vec{\tau}$. The magnetic structure factor for a reflection  $\vec{K}$  is given by:
	\begin{eqnarray}
	\label{magn_str_factor}
	\vec{M}(\vec{K}) &=&{\sum}_j\:\vec{m}_j\:f(\vec{K})\: exp(2i\pi\vec{K}\cdot\vec{r}_j)  \\ 
	  \notag
	\mathrm{with} \;\; \vec{m}_j &=& \frac{1}{2} \cdot\mu(\vec{s}_1-i\vec{s}_2) \:exp(2i\pi \:\vec{\tau}\cdot\vec{r}_j)\: exp(i\phi_j)
	\end{eqnarray}
where $\mu$ is the amplitude of the Mn moments, $f(\vec{K})$  the magnetic form factor of the Mn atoms, $\vec{r}_j$ an atomic position, and $\phi_j$  an arbitrary phase angle. As the magnetic interaction vector  $\vec{M}_\perp(\vec{K}) =\vec{K}\times\vec{M}(\vec{K})\times\vec{K}$ is the projection of the magnetic structure factor onto a plane perpendicular to the scattering vector $\vec{Q}=\vec{K}$, it is convenient to choose the set of orthogonal polarization axes with:\\
\hspace*{15pt} $\cdot \; \hat{x}  \; \| \; \hat{Q}$, \\ 
\hspace*{15pt} $\cdot \; \hat{z}$  perpendicular to the scattering plane and \\ 
\hspace*{15pt} $\cdot \; \hat{y}$  completing the right-handed cartesian set. \\ 
 In the geometry of the present experiment  $\hat{y}$=$\hat{n}$ with $\hat{n}$  the propagation vector of the incoming neutron beam. For the polarimetric measurements the MnSi sample was oriented so that $\vec{Q}=\vec{\tau}_{111}$ and  $\vec{z}\,\|\,(1,\bar{1},0)$. For this reflection the intensity of the diffracted beam was proportional to:
	\begin{equation}
	\label{sigma}
	\sigma =  \vec{M}_\perp\vec{M}_\perp^*\:+\:\vec{P'}\cdot\Im(\vec{M}_\perp \times\vec{M}_\perp^*)
	\end{equation} 					
where $\vec{P'}$  is the polarization of the incoming beam. The second term is the chiral part characteristic of the spiral structure and it adds to the ÒconventionalÓ first term for $\vec{P'}=\hat{Q}$ i.e. when the incident polarization is parallel to $\vec{Q}$. On the other hand, the  reflection is completely extinct  for $\vec{P'}=-\hat{Q}$. Following Blume's equations \cite{Blume}, the scattered polarization for MnSi is given by: 
	\begin{equation}
	\label{polar}
	\vec{P}\sigma=-\vec{P'}(\vec{M}_\perp\vec{M}_\perp^*)+2\Re((\vec{P'}\cdot\vec{M}_\perp^*)\vec{M}_\perp)-\Im(\vec{M}_\perp \times\vec{M}_\perp^*)
	\end{equation} 					
The first two terms form the  `trivial' magnetic part leading to $\vec{P}=-\vec{Q} \, (\vec{Q} \cdot \vec{P'})$ for isotropically distributed electronic magnetic moments\cite{MEZEI_Murani}. The chiral third term creates a polarization antiparallel to $\vec{Q}$  independently from the polarization state of the incoming beam. Both equations \ref{sigma} and \ref{polar} lead to the rigorous and accurate determination of the chiral component $\Im(\vec{M}_\perp \times\vec{M}_\perp^*)$ by polarized neutrons. The more general form of eq.~\ref{polar} is: 
	\begin{equation}
	\vec{P} = \widetilde{P} \vec{P'} + \vec{P^\dagger}
	\label{polar2}
	\end{equation} 					
with $\widetilde{P}$ the polarization transfer tensor and $\vec{P^\dagger}=-\Im(\vec{M}_\perp \times\vec{M}_\perp^*)$ the polarization created by the chiral sample. It is then useful to have the incident polarization along   $-\vec{x}$, $\vec{x}$,  $\vec{y}$  and   $\vec{z}$ and to measure the outgoing polarization components along  $\pm\vec{x}$, $\pm\vec{y}$  and $\pm\vec{z}$ for each configuration of the incident polarization respectively\cite{BrownPNCMI2000}. This measurement procedure is characteristic of spherical neutron polarimetry (SNP) and determines the polarization matrix : 
	\begin{equation}
	\mathbb{P}_{i,j}  =\frac{P'_{i,j}}{| \vec{P}|}  =  \frac{\widetilde{P}_{i,j}\ P'_i \ + P^\dagger_j}{| \vec{P} |}\ \mathrm{with}\ (i,j)\in \{x,y,z\}
	\label{pol_matrix}
	\end{equation} 	
where the denominator ${| \vec{P}|}$ corrects for the finite efficiency of the neutron polarizer-analyzer system and the intrinsic imperfections of the experimental setup.  For nuclear (coherent) scattering  $\mathbb{P}_{i,j} =\delta_{ij}$, i.e. 1 for $i=j$ and 0 otherwise: 
	\begin{equation*}
	\mathbb{P}_{nuclear}\ = \begin{vmatrix}1 &  0 &  0 \\  0 & 1 &  0 \\   0 &  0 &  1\end{vmatrix}
	\end{equation*} 		
For an ideal paramagnet all matrix elements are zero except $\mathbb{P}_{xx} = -1$: 
	\begin{equation*}
	\mathbb{P}_{para}\ = \begin{vmatrix}-1 &  0 &  0 \\  \ 0 &  0 & 0 \\  \ 0 & 0 & 0\end{vmatrix}
	\end{equation*} 		
In MnSi according to eq.~\ref{polar} the scattered beam will always have a polarization antiparallel to the helix propagation vector, i.e. antiparallel to $\hat{x}$ and $\hat{Q}$. For this reason the first row of the matrix will be non-zero, ideally $\mathbb{P}_{i,x} = -1$,  $\mathbb{P}_{i,y\, or\, z} = 0$  with  $(i)\in \{x,y,z\}$. In its most general form the chiral matrix can be written as:
	\begin{equation*}
	\mathbb{P}_{chiral}\  = \begin{vmatrix}-1 & \eta\,\zeta & \eta\,\zeta \\  \ 0 &\ 0 & \ 0 \\  \ 0 & \ 0 & \ 0\end{vmatrix}
	\label{chiral_matrix}
	\end{equation*} 					
where $\zeta$ determines the chirality of the helix: $\zeta=+1$ for right and $\zeta=-1$ for left handed chirality. $\eta$ measures  the fraction of the dominant chiral domain: $\eta$ = 1 for a single domain and $\eta$ = 0 for equally populated  chiral domains or for the disordered paramagnetic state, in which case 
$\mathbb{P}_{chiral}$ reduces to $\mathbb{P}_{para}$. 
\\ \indent
The  polarization matrix of MnSi below  T$_C$  is that of an ideal chiral left handed single domain structure. At T$_C$-4\;K we found:
	\begin{multline*}
	\mathbb{P}_{\mathbb{T}_C-4 K} =  \\
	= \begin{vmatrix}-1.000\pm 0.001 & -0.995 \pm 0.001 & -1.002 \pm 0.001 \\
	-0.007 \pm 0. 001 & 0.016 \pm 0.002 &  -0.007 \pm 0.002 \\
	\;\;\; 0.054 \pm 0.003 &  0.055 \pm 0.002 &  \;\; 0.062 \pm 0.002
	\end{vmatrix}
	\end{multline*} 	
Just above T$_C$, the intensity at $\vec{\tau}_{111}$ drops dramatically, which leads to higher counting times and error bars. Nevertheless, the matrix remains unaffected and is that of a perfect left handed single chiral domain up to $\sim$T$_C$+1, e.g. at T$_C$+0.4 K:  
	\begin{equation*}
	\mathbb{P}_{\mathbb{T}_C+0.4 K}  = \begin{vmatrix}
	 -1.03 \pm 0.03 &  -0.99 \pm 0.03 &  -0.98 \pm 0.03 \\
	-0.01 \pm 0.05 & -0.00 \pm 0.05 &  -0.03 \pm 0.05 \\
	\;\; 0.07  \pm 0.05 &  \ \;\;0.04 \pm 0.01 &  \  \;\; 0.07 \pm 0. 05
	\end{vmatrix}
	\end{equation*} 	
Above  $\sim$T$_C$+1 the non-diagonal elements decrease slowly and chirality remains finite even well above T$_C$. E.g. at T$_C$+4 K we found:
	\begin{equation*}
	\mathbb{P}_{\mathbb{T}_C+4 K}  = \begin{vmatrix} -1.02\pm 0.02 &  -0.4 \pm 0.1 &  -0.3 \pm 0.1 \\
	-0.01 \pm 0.1  & -0.08 \pm 0.1 & -0.05 \pm 0.1\\
	\;\; 0.02 \pm 0.1 & \ \;\;0.07 \pm 0.1 & -0.08 \pm 0.1 
	\end{vmatrix}
	\end{equation*} 					
All data have been corrected for the background determined from polarimetric measurements at 50 K. 
Fig.~\ref{chiral_fraction} shows that $\mathbb{P}_{xy}$  and $\mathbb{P}_{xz}$ are within the error bars unaffected by T$_C$ and start to decrease only above $\sim$T$_C$+1 K.  This result is confirmed by the intensity of the reflection when $\vec{P'} \, \| \,\pm\vec{x}$, in which case eq.~\ref{sigma} becomes: 
	\begin{eqnarray}
	\label{sigma2}
	\sigma &=&  N_{para}\:\pm \; N_{chiral} \;\;\;\;\; \mathrm{with}\\
	N_{para} &=& \vec{M}_\perp\vec{M}_\perp^* \;\;  \mathrm{and} \;\; N_{chiral}= |\vec{P'}\cdot \vec{P^\dagger}|  \nonumber
	\end{eqnarray} 					
where the sign in front of the chiral contribution, $N_{chiral}$, depends on the direction of  $\vec{P'}$  (parallel or antiparallel) with respect to $\vec{Q}$. The intensities in the spin flip channel are: 
\begin{align}
\label{intensities}
N_{x, -x}\equiv  N^{SF}_{\vec{P'}\;\parallel \; Q} = N^{SF}_{bck} + N_{para} + N_{chiral} \\ 
 \notag
N_{-x, x}\equiv N^{SF}_{\vec{P'}\parallel -Q} =N^{SF}_{bck} + N_{para} - N_{chiral}
\end{align} 					
 The chiral fraction $N_{chiral}/( N_{chiral}+ N_{para})$  can then be  deduced from the ratio 
($N_{x, -x}-N_{-x,x})/(N_{x,-x}+N_{-x,x}-2*N^{SF}_{bck})$, with the background $N^{SF}_{bck}$ determined at 50 K. This method minimizes the error bars and systematic errors because both  $N_{x,-x}$ and  $N_{-x,x}$ are measured under identical conditions. The deduced chiral fraction is also shown in fig.~\ref{chiral_fraction} and  coincides with the non-diagonal matrix elements. 
\\ \indent
Thus MnSi remains completely single domain chiral up to T$_{C}'\sim$T$_C$+1 K, i.e. in the complete temperature range, where  the isotropic scattering at the surface of the sphere with radius $\tau$ is observed. We identify T$_{C}'$ as the temperature, where the theoretically predicted skyrmion spin liquid phase sets in, and the analysis of the dynamic and static correlations in the following sections will confirm this hypothesis. 
\section{Neutron Spin Echo}\label{NSE}
\begin{figure}
			\includegraphics[width=85mm]{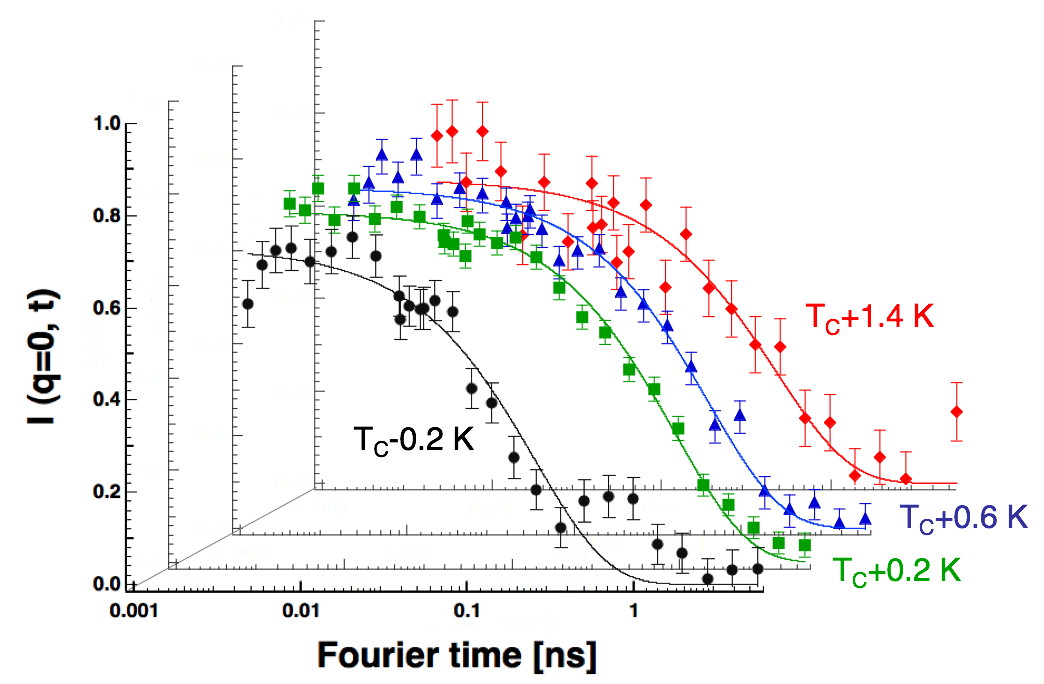}
			\caption{Dynamic correlations at $\vec{\tau}_{110}$, where the spectra are not contaminated by the helical Bragg peaks. }
			\label{NSE_110}
			\end{figure}

\begin{figure}
			\includegraphics[width=85mm]{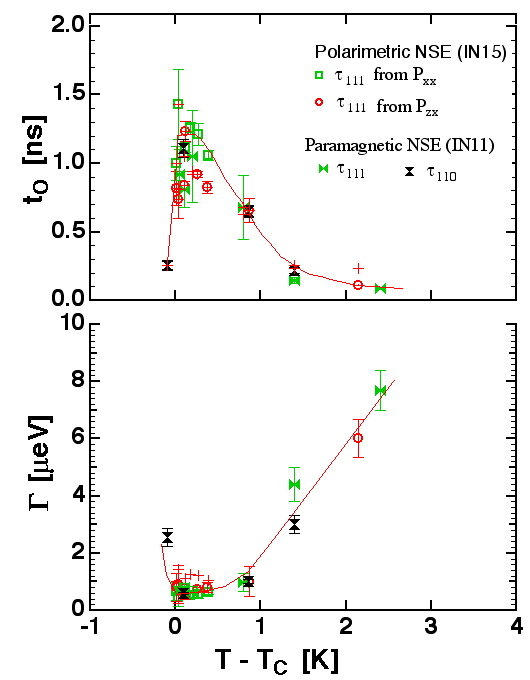}
			\caption{Temperature dependence of the characteristic times t$_0$ (a) and the deduced Lorentz linewidths (b) of the fluctuations at $\vec{\tau}_{111}$ and $\vec{\tau}_{110}$ measured by classical (paramagnetic) and polarimetric NSE. The lines are guides to the eye. }
			\label{gamma_temp}
			\end{figure}

 \indent
The dynamics of MnSi was thoroughly investigated by triple axis spectroscopy (TAS) in the mid-80s \cite{Ishikawa}. It was shown that  strong magnetic correlations exist up to room temperature and that the magnetic excitation spectrum is well described by the Moriya-Kawabata theory for weak itinerant ferromagnets. The TAS measurements, however, did not have the resolution in energy transfer required to analyse the magnetic fluctuations close to T$_C$. With Neutron Spin Echo spectroscopy we can reach the required resolution, follow very accurately the slowing down of the magnetic fluctuations and complement the TAS experiments.  The fact that MnSi is completely chiral above T$_C$ was an unexpected outcome of the polarimetric measurements and also a posteriori the justification for combining NSE and Cryopad to polarimetric NSE.
			\begin{figure}
			\includegraphics[width=85mm]{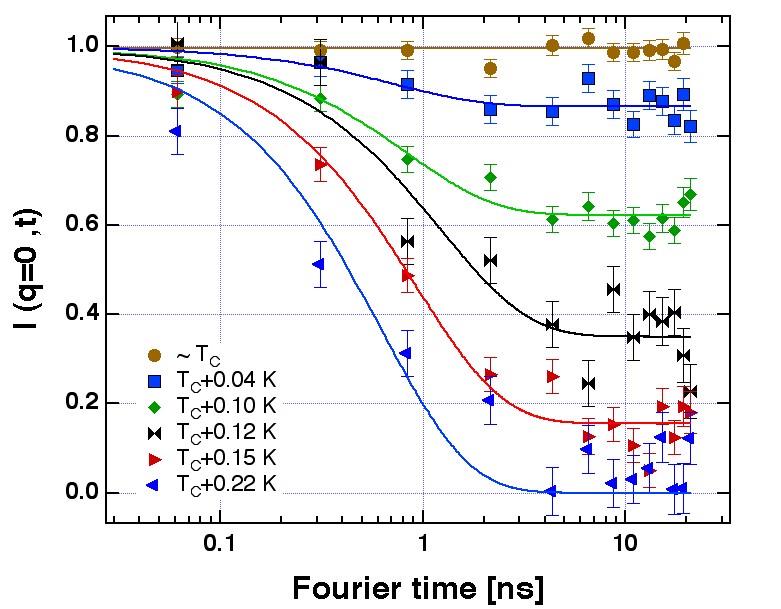}
			\caption{Dynamic correlations at the position of one of the helical Bragg peaks $\vec{\tau}_{111}$ measured by polarimetric NSE for the $\mathbb{P}_{z,x}$ term of the polarization matrix (eq.~\ref{pol_matrix}).  }
			\label{NSE_IN15}
			\end{figure}
			\begin{figure}
			\includegraphics[width=85mm]{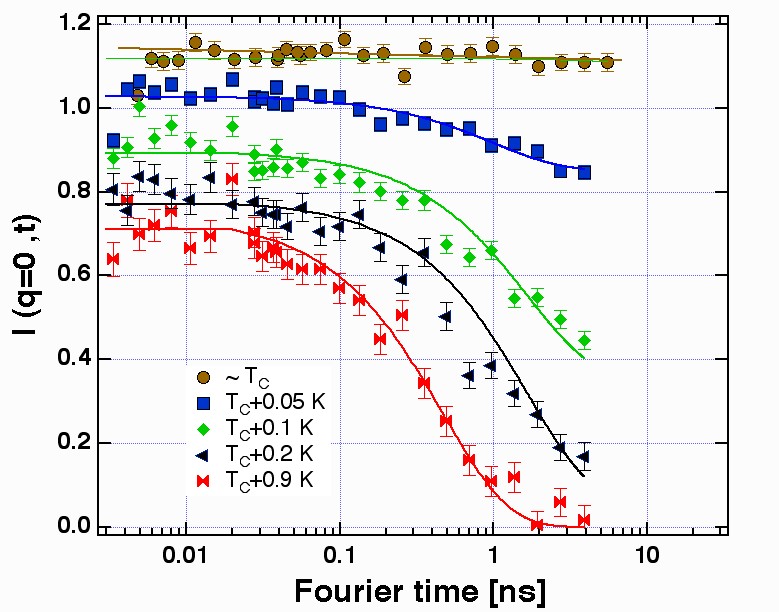}
			\caption{Dynamic correlations at  $\vec{\tau}_{111}$ measured by classical NSE. }
			\label{NSE_IN11}
			\end{figure}
\\ \indent
With the Cryopad on IN15 we were able to measure separately the relaxation of the diagonal  $\mathbb{P}_{xx}$ and of the crossed chiral terms  $\mathbb{P}_{yx}$  and $\mathbb{P}_{zx}$. The corresponding intermediate scattering functions I(q,t) were obtained by normalizing the spectra against the resolution measured below T$_C$, typically at 25 K. Fig.~\ref{NSE_IN15} shows polarimetric NSE spectra for the crossed term  $\mathbb{P}_{yx}$  at $\vec{\tau}_{111}$.  This was the first time that magnetic chiral fluctuations were measured directly and independently from the ÒtrivialÓ diagonal ($\mathbb{P}_{xx}$) part.
\\ \indent
The polarimetric NSE spectra display a purely exponential relaxation superimposed on an elastic term:
	\begin{equation}
	I(q,t)=a \;\exp(-t/t_0) + (1-a)
	\label{NSE_exp}
	\end{equation}
As seen in fig.~\ref{NSE_IN15}  the elastic fraction $(1-a)$  evolves from $\sim$15\% to 100\% within 0.15 K following the fast increase of the intensity displayed in fig.~\ref{intensity} and masking the dynamics at T$_C$. This behaviour is confirmed by standard NSE measurements (fig.~\ref{NSE_IN11}), which however do not extrapolate to 1 at t$\rightarrow$0. The comparison with polarimetric NSE identifies this as an artefact due to the chiral magnetic scattering and not to additional dynamic components. 
\\ \indent
At $\vec{\tau}_{110}$ the quasi-elastic scattering does not develop to a Bragg peak and the NSE spectra were not contaminated by an elastic contribution. Consequently the fluctuations could be followed even below T$_C$ (fig.~\ref{NSE_110}). The decay is exponential with the same  $t_0$ and deduced $\Gamma$ as at $\vec{\tau}_{111}$. 
\\ \indent
All relaxation times and linewidths at $q=0$ are plotted against temperature in fig.~\ref{gamma_temp}. The diagonal ($xx$) and crossed terms ($yx$ and $zx$) measured in the polarimetric NSE mode are also included. All data, even those in the closest vicinity of T$_C$, fall on the same curve. Consequently, the relevant parameter for the fluctuations is the distance from the sphere with radius $\tau$, not the Bragg peaks $\vec{\tau}_{111}$ of the helical phase. 
			\begin{figure}
			\includegraphics[width=85mm]{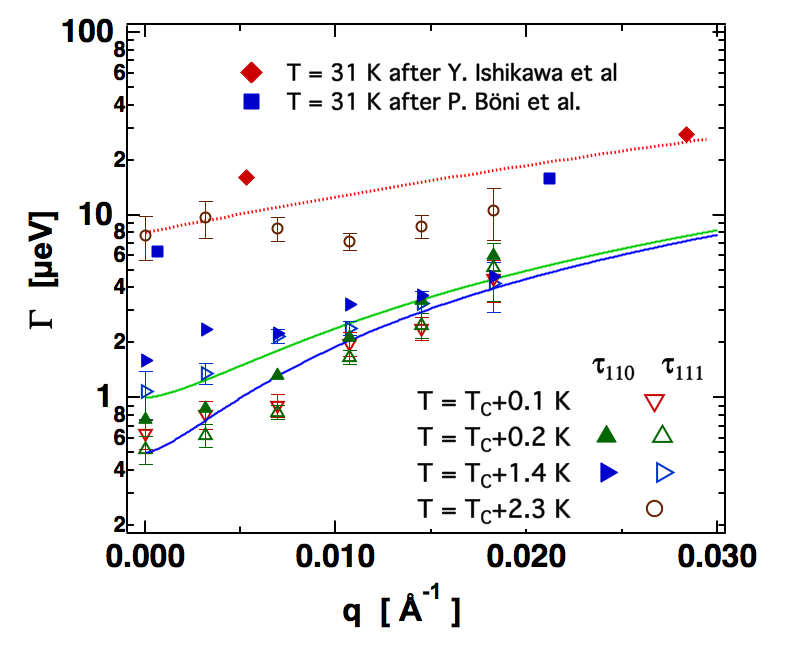}
			\caption{$q$ dependence of the dynamic linewidth $\Gamma$. The open and closed symbols represent the values at $\vec{\tau}_{111}$ and $\vec{\tau}_{110}$ respectively. At the highest temperature ($\sim$ 31 K) there is excellent agreement with literature\cite{Ishikawa, Haussler} and eq. \ref{scaling_Ishi_2} (red dotted line).  }
			\label{gamma_plot_q}
			\end{figure}
\\ \indent
We will now discuss the $q$-dependence of the relaxation. Close to T$_C$ the linewidths take their lowest values at the surface of the sphere with radius $\tau$. We found that $\Gamma$ increases both for $|\vec{Q}|>\tau$ and  $|\vec{Q}|<\tau$ and the low-q points in fig.~\ref{gamma_plot_q} are the average for q=$\pm$0.03 \AA~$^{-1}$. However, above $\sim$T$_C$+1.4 K the q-dependence of $\Gamma$ flattens and it is no longer possible to identify the position of the minimum. 
\\ \indent
Close to T$_C$ $\Gamma(q=0)$ levels off at $\sim$ 0.64~$\mu$eV, which implies that the associated correlation length does not diverge at T$_C$. The results will be discussed in the frame of dynamic scaling after the determination of the correlation length in the following section.
			\begin{figure}
			\includegraphics[width=85mm]{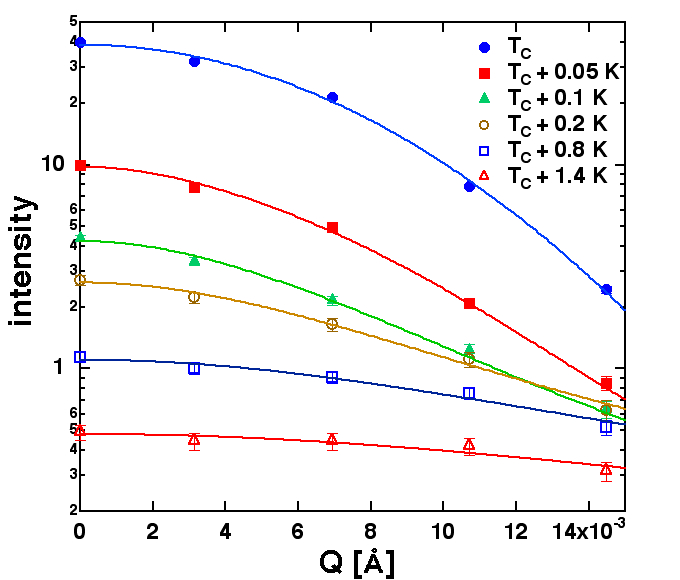}
			\caption{Magnetic signal, proportional to $S(q)$, around the $\vec{\tau}_{111}$. The continuous lines are the best fits: simple Gauss function at T$_C$, superposition of a fluctuating Lorentz (Orstein-Zernike) and an elastic Gaussian part at T$_C<\,$T$\,<\,$T$_C+0.2$~K and a fluctuating Lorentz function above T$_C+0.2$~K. }
			\label{SQ_111}
			\end{figure}
\section{Correlation Length} \label{xi}
			\begin{figure}
			\includegraphics[width=85mm]{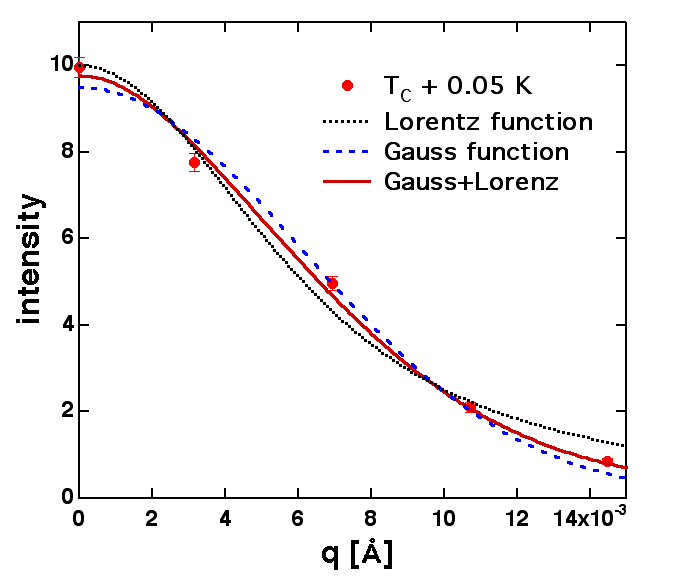}
			\caption{Comparison of the quality of fits for a simple  a simple Lorentz (black dotted line), Gauss (blue dashed line), and the superposition of a Lorentz and a Gauss (red continuous line) at T$_C$+0.05 K and $\vec{\tau}_{111}$  }
			\label{Gauss_lorentz}
			\end{figure}
			\begin{figure}
			\includegraphics[width=85mm]{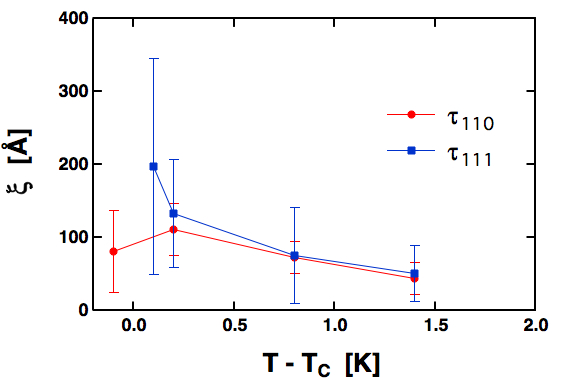}
			\caption{Plot of the correlation lengths ($\xi_{111}$ and $\xi_{110}$) measured at $\vec{\tau}_{111}$ and $\vec{\tau}_{110}$ as a function of temperature.  }
			\label{ksi_IN11}
			\end{figure}
			\begin{figure}
			\includegraphics[width=85mm]{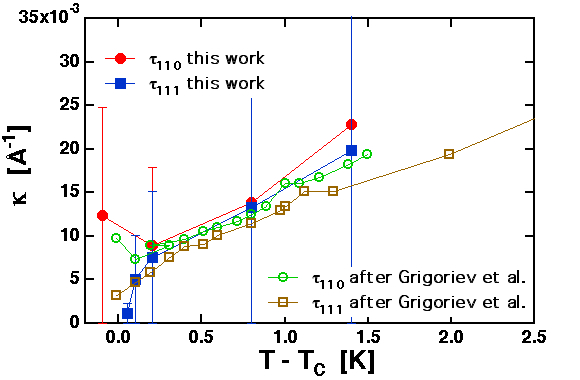}
			\caption{Plot of $\kappa$ measured at  $\vec{\tau}_{111}$ and $\vec{\tau}_{110}$. The closed symbols are the present work. The open symbols are from Grigoriev et al.\cite{Grigoriev}.  }
			\label{kappa_all}
			\end{figure}
			\begin{figure}
			\includegraphics[width=85mm]{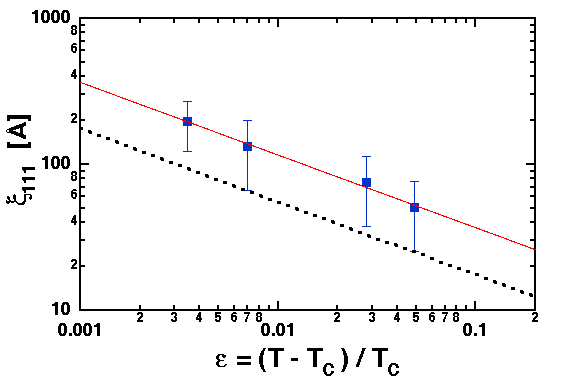}
			\caption{Plot of the correlation legth $\xi_{111}$ determined at  $\vec{\tau}_{111}$ versus the reduced temperature $\epsilon$ on a log-log scale. The continuous line is the best power law fit $\xi= 12\; \epsilon^{-0.5}\;$\AA . The black dotted line shows the extrapolated values from the high temperature TAS work of Ishikawa et al.\cite{Ishikawa}. }
			\label{ksi_log}
			\end{figure}
\indent
The $Q$ dependence of the magnetic static structure factor $S(Q)$ was analysed in the configuration $N_{x, -x}$ of eq.~\ref{intensities}, where the  magnetic intensity is maximum and the background correction negligible for T$<$30 K.   
\\ \indent
Fig.~\ref{SQ_111} shows the scattered neutron intensity, which is proportional to $S(Q)$, around $\vec{\tau}_{111}$ in a log-lin scale. At T$_C$ the points are best fitted with by a Gauss function, with $\sigma^2=6.11\, 10^{-3}\, \pm 3\,10^{-5}$\AA $^{-1}$  leading to $1.44\, 10^{-2}\pm10^{-4}$ \AA $^{-1}$  FWHM, i.e. significantly broader than the $Q$-resolution of the instrument. Well above T$_C$, the data are well described by the OZ function of eq.~\ref{static} convoluted with the Gauss resolution function.
\\ \indent
Just above T$_C$, however, neither the Gauss nor the OZ functions fit satisfactorily the experimental data. Instead, the best fit is obtained by a superposition of a fluctuating OZ and an elastic Gauss with relative weights fixed from the NSE spectra, as shown by fig.~\ref{Gauss_lorentz}. We note that all three fits in fig.~\ref{Gauss_lorentz} involve the same number (two) of independent parameters:  the total intensity and the $\sigma$ for the Gauss or $\kappa$ for the Lorentz and Gauss+Lorentz fits respectively (in this case the width of the Gauss was fixed to that found below T$_C$). 
\\ \indent
Similarly to the NSE spectra, the data analysis is easier at $\vec{\tau}_{110}$, where there is no contamination from Bragg peaks. At this position of the reciprocal space a simple Lorentzian describes the scattered neutron intensity at all temperatures, even below T$_C$. All correlation lengths are plotted varsus T-T$_C$ in fig.~\ref{ksi_IN11} and the  deduced values of $\kappa$ are in excellent agreement with previous published data\cite{Grigoriev} (fig.~\ref{kappa_all}).  
\\ \indent
At $\vec{\tau}_{110}$  the correlation length $\xi_{110}$ levels off at about 100~\AA~, i.e. about half the pitch of the helix. At $\vec{\tau}_{111}$  the correlation length $\xi_{111}$ increases considerably close to T$_C$ following a power law of the reduced temperature $\epsilon$ similar to those found at second order phase transitions and illustrated by fig.~\ref{ksi_log}. The continuous line in the figure corresponds to:
	\begin{equation}
	\xi_{111} = a \; \epsilon ^{-\nu}\;\; \mathrm{with} \;\; a=12 \pm 2 \, \mathrm{\AA} \; \mathrm{and}\;\; \nu=0.5 \pm 0.2
	\label{ksi_power_law}
	\end{equation}
The black dotted line represents  $\xi= 5.6 \; \epsilon ^{-0.5}$ \AA, the extrapolated curve from the high temperature TAS data of Ishikawa et al.\cite{Ishikawa}. It is remarkable that this extrapolation from very high temperatures (100 K - 300 K ) is only a factor 2 off our results. 
			\begin{figure}
			\includegraphics[width=85mm]{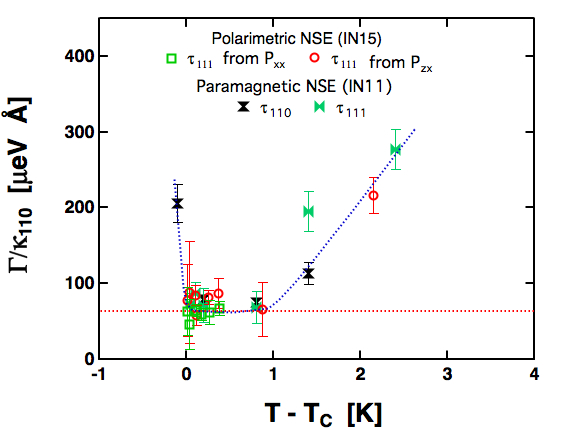}
			\caption{Temperature dependence of the ratio $\Gamma$(q=0)/$\kappa_{110}$ revealing the existence a broad flat  minimum, where $\Gamma$(q=0) $\propto\; \kappa_{110}$ between T$_C$ and T$_{C}'$. }
			\label{tau_ksi_temp}
			\end{figure}
\section{Dynamic Scaling} \label{scaling}
\indent
Dynamic scaling relates $\Gamma$ to $\kappa$ (or inversely $t_0$  to $\xi$) through a homogenous function: 
	\begin{equation}
	\Gamma(q, \epsilon)  \propto \Gamma(q, \epsilon=0)  \, f (\kappa/q) \\ 
	\label{dynamic_scaling}
	\end{equation}
with the dimensionless ratio $\kappa/q$ defining the critical ($\kappa/q \ll 1$) and hydrodynamic ($\kappa/q \gg1$) regimes respectively\cite{Halperin}. At the critical limit, which is always reached  at $\kappa$=0, the linewidth and relaxation times reflect the volume probed by the measurement :
	\begin{subeqnarray}
	 \mathrm{t}_0(q, \epsilon=0) &\propto& \, q^{-z} \;\;  \mathrm{and} \\
	 \Gamma(q, \epsilon=0) &\propto& \, q^z \
	 \label{dynamic_power_law}
	\end{subeqnarray}
with $z$ the dynamic exponent. For $q = Q$ as in ferromagnets $z=5/2$, whereas for antiferromagnets and MnSi, where the relevant parameter is $q=|Q-\tau|$, $z=3/2$.
			\begin{figure}
			\includegraphics[width=85mm]{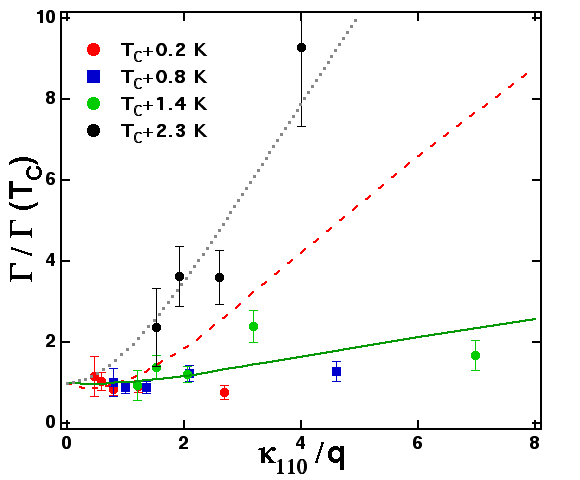}
			\caption{Scaling plot of the reduced dimensionless quantities $\Gamma/\Gamma($T$_C)$ versus  $\kappa_{110}/q$. The black dotted line shows the extrapolated values from the previous high temperature TAS data (eq. \ref{scaling_Ishi_2}). The red dashed line is the Resibois-Piette function for antiferromagnets and the green continuous line is a guide to the eyes through the experimental data for T$_C  \lesssim $T $\lesssim$T$_C'$.}
			\label{scaling_2}
			\end{figure}
\\ \indent
At the hydrodynamic limit, $q=0$ and $\kappa > 0$, the dynamics is governed by the correlated volumes through the power laws: 
	\begin{subeqnarray} \label{z}
	 \;\;t_0(q=0) & \propto & \xi^z \slabel{xi_z} \;\; \mathrm{and\;}\\
	 \Gamma(q=0)  &\propto& \kappa^z  \slabel{kappa_z}
	\end{subeqnarray}
\indent
The previous extensive study of spin fluctuations in MnSi with TAS\cite{Ishikawa} showed that the linewidth is equally well described by two universal scaling functions over an extremely large temperature scale, from $\sim$30~K up to the room temperature: 
	\begin{subeqnarray} \label{scaling_Ishi}
	  \Gamma  &\sim& \;\; \Gamma_0\, Q^2 \;\; [ 1+(\kappa/Q)^2]  \slabel{scaling_Ishi_1} \\
	  &\sim&  \mathrm{A}_0 \,Q^{5/2} \;  [  1+(\kappa/Q)^2]  \slabel{scaling_Ishi_2}
	\end{subeqnarray}
with  $\kappa^2 = 0.0325 \, \epsilon$ \AA$^{-2}$, $\Gamma_0$=50~meV\AA$^3$ and A$_0$=19.6~mev\AA$^{5/2}$. Equation~\ref{scaling_Ishi_1} is derived from the Moriya-Kawabata theory for weak itinerant ferromagnets\cite{Moriya_Kawabata} whereas eq.~\ref{scaling_Ishi_2} has the form expected for dynamic scaling. The red dotted line going through the data at T$_C$+2.3~K in fig~\ref{gamma_plot_q} corresponds to the calculated values from eq.~\ref{scaling_Ishi_2} and is in excellent agreement with our experimental results.  
 \\ \indent
At high temperatures, the momentum transfer $Q$ is the relevant parameter and $z=5/2$. Close to T$_C$, however, magnetic correlations build up at $\tau$ and the relevant parameter crosses over to $q=|Q-\tau|$, which is used in this work. Consequently also the dynamic critical exponent should cross over from $z=5/2$ to $z=3/2$.  
\\ \indent
 If the transition were of second order, an analysis along the lines of dynamic scaling would imply critical slowing down. In this case $\Gamma_{111}(q=0)$ should decrease continuously to zero at T$_C$ following the divergence of $\xi_{111}$. In contrast, $\Gamma_{111}(q=0)$ remains finite and strictly the same as  $\Gamma_{110}(q=0)$. Moreover, fig.~\ref{gamma_temp} and fig.~\ref{kappa_all} show that close to T$_C$ both $\Gamma_{111}$ and $\Gamma_{110}$ are roughly proportional to $\kappa_{110}$,  not to the ``critical'' $\kappa_{111}$. Consequently, we do not observe the critical slowing down and rapidly changing dynamic behaviour expected for second order phase transitions. This is underlined by  fig.~\ref{tau_ksi_temp} where the ratio between all $\Gamma$ and the calculated $\kappa_{110}$ levels off to a broad plateau between T$_C$ and T$_C'$, pointing towards $z\sim1$.
 \\ \indent
Fig.~\ref{scaling_2} shows the interdependence of the reduced dimensionless quantities $\Gamma/\Gamma($T$_C)$ and $\kappa_{110}$/q. 
Following the scaling assumption of eq.~\ref{dynamic_scaling} all experimental data should collapse on a universal curve. This is obviously not the case. Close to T$_C$ both $\Gamma$ and $\kappa_{110}$ do not vary with temperature and for this reason the data accumulate on the green continuous line of fig.~\ref{scaling_2}, which is  significantly different from the Resibois-Piette function for antiferromagnets\cite{resibois_piette}. Above $\sim$T$_C'$ there is a fast change of behaviour crossing over to eq.\ref{scaling_Ishi} and the black dotted line in fig.~\ref{scaling_2}. 
 \\ \indent
\section{Discussion} \label{discussion}
			\begin{figure}
			\includegraphics[width=85mm]{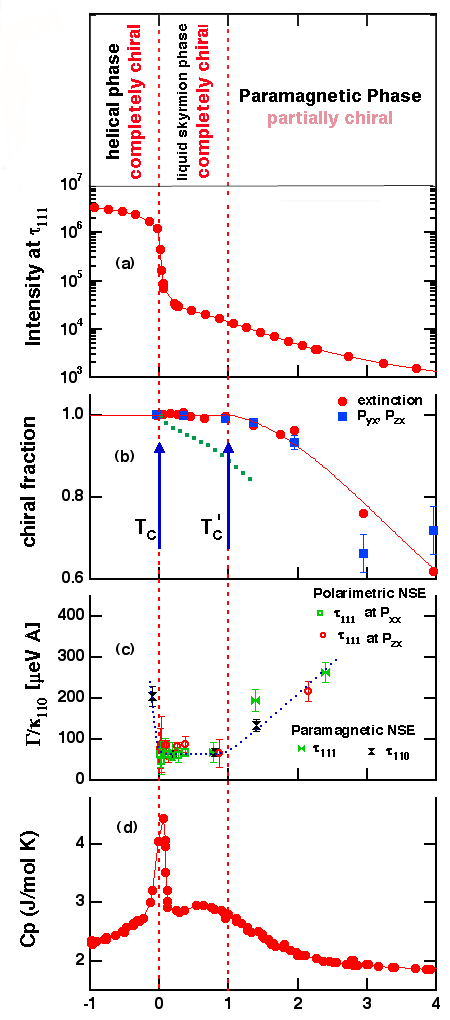}
			\caption{Temperature dependence of (a) the intensity at the helical Bragg peak $\vec{\tau}_{111}$  (b) the chiral fraction, (c) the ratio $\Gamma/\kappa_{110}$ and  (d) the specific heat from Stishov et al. \cite{StishovPRB}. These experimental results confirm the  existence of a new completely chiral but fluctuating skyrmion liquid phase between T$_C$ and T$_C'$.} 
			\label{final_all}
			\end{figure}
These high resolution neutron scattering data confirm  the first order nature of the helical transition in MnSi seen by specific heat and ultrasonic attenuation measurements\cite{StishovPRB, StishovJPhysC}. In favour of the first order phase transition are: 
\\ \indent
a. the sharpness of the transition (fig.~\ref{intensity}), even though the helical Bragg peaks are not resolution limited. 
\\ \indent
b. the coexistence of high and low temperature phases between T$_C$ and ~T$_C$+0.2 K. At $\vec{\tau}_{111}$ the NSE spectra are the superposition of a fluctuating and elastic part (fig.~\ref{NSE_IN15} and \ref{NSE_IN11}) and the static structure factor is best described by the superposition of a Lorentzian (fluctuating) and Gaussian (elastic) contributions (fig.~\ref{Gauss_lorentz}).
\\ \indent
c. the absence of critical slowing down at T$_C$. The slow change of the dynamics scales with the non diverging correlation length $\xi_{110}$.
\\ \indent
Fluctuations are present above T$_C$ and  $\xi_{111}$ can indeed be approximated by a power law (eq.~\ref{ksi_power_law} and fig.~\ref{ksi_log}). For this reason, a selection of experimental data can always be bent to fit into the second order phase transition scheme by introducing crossovers, as it was recently done by Grigoriev at al.\cite{grigoriev_prb2}. However, second order phase transitions are global transformations involving all parameters of the system, and this is not the case in MnSi. 
\\ \indent
We will now proceed to the discussion of the fluctuating phase between T$_C$ and T$_C'$. Fig.~\ref{final_all} recapitulates  the most important findings of this work including the specific heat measured by Stishov et al.\cite{StishovPRB}. The first order helical transition T$_C$ is marked by the sharp peak in the specific heat (fig.~\ref{final_all}d) and the Bragg peak (fig.~\ref{final_all}a). However,  the helical transition has no effect on $\eta$, the degree of single domain chirality. Comparison with theory shows that this is not a trivial effect. $\eta$ can indeed be calculated in the frame of a simple mean field model\cite{Grigoriev} :
 \begin{equation}
\eta=2Q\tau/(Q^2+\tau^2+1/\xi^2)
\label{eta}
\end{equation} 					
The measured $\xi_{111}$  lead to the dotted green line of fig.~\ref{final_all}b, which is  significantly different from the  experimental data. Consequently, the simple assumption of  an unpinned and fluctuating helical phase is not confirmed by the experiment. In contrast,  $\eta$=1 implies a high degree of chiral short range order for the fluctuating phase between  T$_C$ and T$_{C}'$. 
\\ \indent
T$_{C}'$ is exactly the theoretically predicted temperature for the formation of a skyrmionic ground state in MnSi\cite{Skyrmion}. Fig.~\ref{final_all} shows that this is not a singular event but part of a global transformation involving the dynamics as well as a broad maximum in the specific heat\cite{StishovPRB}. Besides specific heat,  thermal expansion, sound velocity, sound absorption as well as  resistivity measurements point towards a non trivial spin liquid phase \cite{Petrova}.  Skyrmions account for the inherent topology of this completely chiral spin liquid  phase, which gives a scattering with rotational symmetry. Skyrmionic textures are energetically favourable at short distances, which explains the slow decrease of $\eta$ and its finite value well above T$_C$. The correlation length $\xi_{110}$ gives a natural measurement of the skyrmion radius, which peaks are about half of the helix pitch close to T$_C$, in agreement with the skyrmion hypothesis. 
\\ \indent
The Bragg peak of fig.~\ref{final_all}a is a direct measurement of the helical order parameter. We speculate that $\eta$ in fig.~\ref{final_all}b reflects the order parameter of the  skyrmion spin liquid phase. The topology of skyrmions would explain the  particular dynamics and the low dynamic exponent of $z\sim1$. This skyrmion liquid phase would then be the magnetic equivalent of the blue phases in liquid crystals, which occur at a very limited temperature range above the helical phase. In this way the phase diagram of MnSi becomes very similar to that of cholesteric liquid crystals, emphasizing the parallel between magnetic and structural ground states. Similarly to liquid crystals, the transition to the helical state is of first order due to the different topologies of the helical and skyrmionic phases.
\\ \indent
In summary, we combined  high resolution neutron spin echo spectroscopy and spherical polarimetry to obtain a consistent picture of the spin liquid phase of MnSi above T$_C$. The results evidence a first order transition between the helical phase and a  skyrmion liquid phase. This  completely (single domain)  chiral and strongly fluctuating new state of matter occurs in a narrow temperature range above the helical phase of MnSi. 
 

\acknowledgments
C.P. thanks  P. B\"{o}ni, U. R\"{o}ssler and  E.L.-B. thanks P.J. Brown for  fruitful discussions. The authors acknowledge  the support of the ILL technical teams in particular E. Bourgeat-Lami, C. Gomez and E. Thaveron. Special thanks go to Thomas Krist for the compact solid state polarizer, which enabled the polarimetric NSE measurements. This project was partly supported by the European Commission under the 6$^{th}$ Framework Programme through the Key Action: Strengthening the European Research Area, Research Infrastructures. Contract no: RII3-CT-2003-505925."

\bibliography{prb_refs}		

\end{document}